\documentclass{ws-ijmpb}

\begin{document}

\catchline{}{}{}{}{}
%
%%%%%%%%%%%%%%%%%%%%%%%%%%%%%%%%%%%%%%%%%%%%%%%%%%%%%%%%%%%%%%%%%%%%

\title{EVOLVING SCALE-FREE NETWORK MODEL WITH TUNABLE CLUSTERING}

\author{BING WANG*$^{\dag}$ ZHONGZHI ZHANG$^{\ddag}$ HUANWEN TANG* ZHILONG XIU$^{\S}$}
\address{*Department of Applied Mathematics, Dalian University of Technology, \\
Dalian, 116023, P. R. China\\}
\address{$^{\ddag}$Institute of System Engineering, Dalian University of Technology, \\
Dalian, 116023, P. R. China\\}
\address{$^{\S}$School of Environmental and Biological Science and Technology, Dalian University of
Technology, Dalian 116023, P. R. China\\}
\address{$^{\dag}$bingbignmath@yahoo.com.cn}
 \maketitle

\begin{abstract}
The Barab\'{a}si-Albert (BA) model is extended to include the
concept of local world and the microscopic event of adding edges.
With probability $p$, we add a new node with $m$ edges which
preferentially link to the nodes presented in the network; with
probability $1-p$, we add $m$ edges among the present nodes. A
node is preferentially selected by its degree to add an edge
randomly among its neighbors. Using continuum theory and rate
equation method we get the analytical expressions of the power-law
degree distribution with exponent $\gamma=3$ and the clustering
coefficient $c(k)\sim k^{-1}+c$. The analytical expressions are in
good agreement with the numerical calculations.
\end{abstract}


\section{Introduction}    %) A SECTION HEADING

Various complex systems both in nature and in society, such as the
Internet,\cite{1} cells \cite{2,3,4} and social networks can all
be described by networks. Despite their diversities, most networks
appearing in nature follow universal organizing principles. In
particular, it has been found that the degree distributions of
most networks decay as a power law, following $p(k)\sim
k^{-\gamma}$, where $\gamma$ is the degree exponent varied between
2 and 3, and the average shortest path length increases with the
system size logarithmically.\cite{5,6,7} Especially for social
networks, which show high clustering for their high transitivity.
That is, if person \emph{A} knows person \emph{B} and \emph{C},
then \emph{B} and \emph{C} are likely to know each other. The
previous studies on social networks have originated from Rapport's
studies of disease spreading.\cite{8} To understand the effect of
network topology on the pattern of disease spread, one must first
understand its topology and evolving mechanisms.

The small-world network model (typical WS model) proposed by Watts
and Strogatz has both the properties of high clustering and a
short path length, while the degree distribution is similar to
Poisson distribution, not coincident with the real- world network
of power-law distribution.\cite{9}  The origin of scale-free
properties is well understood in terms of interaction that
generates this topology dynamically by the Barab\'{a}si-Albert
model, e.g., on the basis of network growth and preferentially
attachment.\cite{10} Although the model captures the power-law
tail of the degree distribution, it has other properties that may
not agree with the empirical results of the real networks, such as
the property of high clustering. In fact, it may be impossible to
construct an universal model to describe all kinds of networks
well.

A great number of attempts have been made to construct a specific
model with some properties to coincide with some special
networks.\cite{11,12,13,14,15,16,17} Davidsen {\it et al}
constructed a small world network with local interaction to model
acquaintance networks. The model is formulated with fixed number
of $N$ nodes. The local connection rule is based on "transitive
linking". By simulation, he got a scale-free network with
parameter of probability $p$ and the network had the small-world
character.\cite{18} Holme {\it et al} constructed an extended
model of scale-free network to include a "triad formation step".
They also obtained a scale-free network with tunable clustering
and a short path length.\cite{19} Then Szab\'{o} {\it et al} gave
the analytical expression of the clustering coefficient varied
with degree $k$ for this model.\cite{20} In this paper, we propose
another extended model to make an attempt to explore the evolving
mechanisms of social networks.

There are many local communities existing in social networks
according to different interests or ages, etc. For example, in a
collaboration network of scientists at the Santa Fe Institute,
scientists are classified into different communities according to
their similarity either of research topic or methodology. The
communities include scientists whose research are on the structure
of RNA, statistical physics and mathematical ecology.\cite{21}
Another character of the social network is its high transitivity.
It is more reasonable to consider that the center person is more
likely to make his neighbors acquainted by introducing them to
each other than a person selected randomly. Based on these two
assumptions we propose an evolving model of scale-free network
with tunable clustering.

The rest of the paper is organized as follows: Sec.2 provides a
brief review of the Barab\'{a}si-Albert scale-free network
generation algorithm, then our model and algorithm is introduced,
followed by analytical calculations with continuum theory and rate
equations method in Sec.3. Finally, some conclusions are given in
Sec.4.

\section{Models}
\subsection{The Barab\'{a}si-Albert scale-free model}
The BA model is briefly reviewed first and then our algorithm is
proposed. The BA model is defined as follows\cite{10}:

$(i)${\it Initial condition}: The network consists of $m_0$ nodes
and no edges.

$(ii)${\it Growth} : At each time step, add a new node with
$m(m\leq{m_0})$ edges.

$(iii)${\it Preferential attachment}: Each
edge of the new node is attached to the existing node $i$. The
probability $\prod$ that a new node will be connected to node $i$
depends on the degree $k_i$ of node $i$, such that $\prod(k_i)$ is
proportional to its degree $k_i$, that is:

$$\prod (k_i)=\frac{k_i}{\sum\limits_{j} k_j}$$

After $t$ time steps, the system develops to be a network with
$N=m_0+t$ nodes. With continuum theory, Barab\'{a}si and Albert
calculated the degree distribution $p(k)$ analytically. The
probability $p(k)$ that a randomly chosen node has $k$ links
decays as a power law, $p(k)\sim\frac{2m^2}{k^3}$.

\subsection{Our Model}
The BA model can be extended with the concept of local world and
the microscopic event of adding edges. The local-world network was
proposed by Li,\cite{14} which is based on the appearance of local
world in a superfamily of protein in protein interaction network,
human community etc. So the local world concept is also introduced
in our model. The model is generated by the following algorithm:

$(i)${\it Initial condition}: Start with $m_0$ nodes and $e_0$
edges, and select $M$ nodes randomly from the present network,
referred to as the local community of the new coming node.

$(ii)$With probability $p$, add a new node with $m$ edges. The new
node attaches to an existing node $i$ of the local community with
the probability $\prod(k_i)$, which is proportional to its degree.
\begin{equation}
\prod(k_i)=\prod(i\in local community)\cdot\frac{k_i}{\sum_{j\in
local community}k_j}
\end{equation}
where$\prod(i\in local
community)=\frac{M}{N}=\frac{M}{m_0+pt}\approx\frac{M}{pt}$.

$(iii)$With probability $1-p$, add $m$ edges. A node is selected
according to the preferential attachment rule and then an edge is
randomly added among its neighbors. Just as in social networks,
the more famous a person is, the more likely he is to pick two
persons of his neighbors randomly and introduce them to know each
other.

These steps are then iterated. After $t$ time steps, the network
develops to be $ N=m_0+pt$ nodes and $2mt$ edges, then the average
degree of the network is:
\begin{equation}
<k>=\frac{2mt}{m_0+pt}\approx \frac{2m}{p}
\end{equation}

\section{Analytical Results}
The continuum theory and rate equations method are used to
calculate the degree distribution $p(k)$ and the clustering
coefficient $c(k)$ for various degree $k$. According to continuum
theory, $k_i$ is supposed to change continually. The probability
$\prod(k_i)$ is the rate at which $k_i$ changes. For node $i$ of
degree $k_i$, $k_i$ changes as
\begin{equation}
\frac{\partial k_i}{\partial t}=pm\prod(k_i)+(1-p)m\prod(i\in
local community)\sum_{n\in\Omega }\frac{k_n}{\sum_{j\in local
community}k_j}\cdot\frac{1}{k_n}
\end{equation}
where $\Omega$ is the neighbors set of node $i$.
 The first term in the sum corresponds to node $i$ selected
 preferentially to link to the new node in step $(ii)$; the second term is own to step $(iii)$.
 Node $i$ is one of the neighbors of preferentially selected node
 $n$, then node $i$ is randomly selected to add an edge to increase the degree of node $i$.
$k_n$ denotes the degree of the neighboring node $n$ of node $i$.

 Using(2), we can obtain:
\begin{eqnarray}
\frac{\partial k_i}{\partial t}&=&pm\prod(i\in local
 community)\cdot{\frac{k_i}{\sum_{j\in local community}k_j}}\nonumber\\
&&+(1-p)m\cdot\prod(i\in local community)
 \cdot{\frac{k_i\cdot k_n}{\sum_{j\in local community}k_j}}\cdot{\frac{1}{k_n}}\nonumber\\
&\approx&pm\cdot{\frac{M}{pt}}\cdot{\frac{k_i}{M\cdot<k>}}+(1-p)m\cdot{\frac{M}{pt}}\cdot{\frac{k_i}{M\cdot<k>}}\nonumber\\
&\approx& m\cdot{\frac{M}{pt}}\cdot{\frac{k_i}{M\cdot<k>}}\nonumber\\
&\approx&\frac{k_i}{2t}
\end{eqnarray}

 So the change rate of degree $k_i$ is given by:
 \begin{equation}
 \frac{\partial k_i}{\partial t}\approx{\frac{k_i}{2t}}
 \end{equation}
 then
 \begin{equation}
 k_i(t)\approx{m(\frac{t}{t_i})^{1/2}}
 \end{equation}
 With continuum theory, we can also obtain the same degree
 distribution as that of the BA model. The degree distribution follows the
 same power law, $p(k)\sim{\frac{2m^2}{k^3}}$.

 Fig.1 shows that the analytical calculation is in good agreement with numerical
 result.
 \begin{figure}[th]
\centerline{\psfig{file=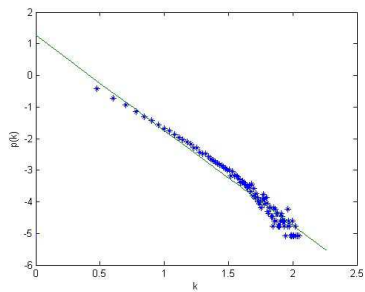,width=9cm}} \vspace*{8pt}
\caption{Degree distribution $p(k)$ in log-log scale with
$N=10000,m_0=25$, $M=20, m=3, p=0.97$. Star-points represent
numerical simulation; Solid line is an analytical expression with
$p(k)\sim{\frac{2m^2}{k^3}}$.}
\end{figure}

To calculate the clustering coefficient of node $i$, the average
rate of change for $e_i$ must be considered, where $e_i$ is the
number of connected neighbors of node $i$. Then by using
$c_i(k_i)=\frac{e_i}{\frac{k_i(k_i-1)}{2}}$, we can get $c_i(k_i)$
for node $i$ of degree $k_i$.

For node $i$, its connected neighbors $e_i$ changes for two cases
as shown in Fig.2. Case I shows the step $(ii)$ to add a new node.
The new node coming into system preferentially attaches to node
$i$ and one of its neighbors to form a triangle; Case II shows the
step $(iii)$ to add an edge. Node $i$ is preferentially selected
to pick two of its neighbors randomly to link. This process also
increases the number of $e_i$, so the rate equation for $e_i$ is
as follows:
\begin{figure}[th]
 \centerline{\psfig{file=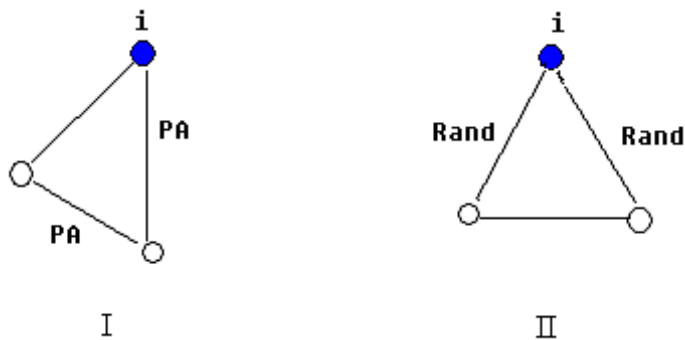,width=9cm}} \vspace*{8pt}
\caption{Two possible cases to increase the number of connected
neighbors $e_i$ of node $i$. Case I: The new node preferentially
attaches to node $i$ and one of its neighbors in step $(ii)$. Case
II: Two neighbors of node $i$ are randomly selected to add an edge
in step $(iii)$.}
\end{figure}

\begin{eqnarray}
I:\frac{\partial e_i}{\partial t}&=&pm\prod(i\in local community)\frac{k_i}{\sum_{j\in local community}k_j} \nonumber\\
&&\cdot(pm-1)\prod(i\in local community)\cdot\sum_{n\in\Omega}
\frac{k_n}{\sum_{j\in local community)}k_j}
\end{eqnarray}

where $\Omega$ is still the neighbors set of node $i$. The right
side of Eq.(7) corresponds to case I in Fig.2. The new node
preferentially links to node $i$ and one of its neighbors. Eq.(7)
is equivalent to the following:
\begin{eqnarray}
I: \frac{\partial e_i}{\partial t}
&=&pm\cdot\frac{M}{m_0+pt}\cdot\frac{k_i}{<k>\cdot M}
\cdot(pm-1)\cdot\frac{M}{m_0+pt}\cdot\frac{k_i\cdot<k_n>}{M\cdot<k>}\nonumber\\
&\approx&\frac{m\cdot k_i}{<k>t}\cdot(pm-1)\cdot \frac{1}{pt}\cdot\frac{k_i\cdot<k_n>}{<k>}\nonumber\\
&\approx&\frac{m(pm-1)}{p}\cdot\frac{k_i^2}{t^2\cdot<k>^2}\cdot<k_n>\nonumber\\
&\approx&\frac{m(pm-1)}{p}\cdot\frac{k_i^2}{t^2\cdot<k>^2}\cdot\frac{p<k>}{4}\cdot
lnN\nonumber\\
&\approx&\frac{m(pm-1)}{4}\cdot\frac{k_i^2}{t^2\cdot <k>}\cdot
lnN\nonumber\\
&\approx&\frac{p(pm-1)}{8}\cdot\frac{k_i^2}{t^2}\cdot lnN
\end{eqnarray}
In Eq.(8), the average degree of the first neighbor of node $i$ is
denoted by $<k_n>$.
For uncorrelated random BA model,
$<k_n>=\frac{p\cdot<k>}{4}\cdot lnN$, see reference \cite{22}.

For part II, we have:
\begin{equation}
II:  \frac{\partial e_i}{\partial t}=(1-p)\cdot m\prod(k_i)
\end{equation}
II is the process of adding edges in Fig.2 case II. Node $i$ is
preferentially selected to add an edge among its neighbors
randomly to increase $e_i$.

Integrating  Eq.(8)and Eq.(9), respectively, we obtain:
\begin{eqnarray}
I:e_i&=&\int_{1}^{\frac{N}{p}}\frac{p(pm-1)}{8}\cdot\frac{k_i^2}{t^2}\cdot ln(pt)dt\nonumber\\
&=&\int_{1}^{\frac{N}{p}}\frac{p(pm-1)}{8}\cdot\frac{m^2t}{t^2\cdot t_i}\cdot ln(pt)dt\nonumber\\
&=&\int_{1}^{\frac{N}{p}}\frac{p(pm-1)}{8t_i}\cdot\frac{m^2}{t}\cdot ln(pt)dt\nonumber\\
&=&\int_{1}^{\frac{N}{p}}\frac{p(pm-1)\cdot m^2}{8t_i}\cdot ln(pt)d(ln(pt))\nonumber\\
&\approx&\frac{p(pm-1)\cdot m^2}{16t_i}\cdot(lnN)^2 \nonumber\\
&\approx&\frac{p^2(pm-1)}{16}\cdot\frac{(lnN)^2}{N}\cdot k_i^2 \nonumber\\
\end{eqnarray}

\begin{eqnarray}
II:e_i&=&\int_{1}^{\frac{N}{p}}(1-p)m\prod(i\in local community)
\cdot\frac{k_i}{\sum_{j\in local community}k_j}dt\nonumber\\
&\approx&\int_{1}^{\frac{N}{p}}(1-p)m\cdot\frac{M}{pt}\cdot\frac{k_i}{M\cdot<k>}dt\nonumber\\
&\approx&\int_{1}^{\frac{N}{p}}\frac{(1-p)}{2}\cdot\frac{k_i}{t}dt\nonumber\\
&\approx&(1-p)\int_{1}^{\frac{N}{p}}\frac{\partial k_i}{\partial t}dt\nonumber\\
&\approx&(1-p)(k_i(\frac{N}{p})-e_0)
\end{eqnarray}

By summing $I$ and $II$, the number of connected neighbors $e_i$
of node $i$ is obtained:
\begin{equation}
e_i=e_{i0}+\frac{p^2(pm-1)}{16}\cdot{\frac{(lnN)^2}{N}}\cdot
k_i^2+(1-p)(k_i(\frac {N}{p})-e_0)
\end{equation}

with $c_i=\frac{e_i}{\frac{k_i(k_i-1)}{2}}$, the general
expression of the clustering coefficient $c_i(k_i)$ has the form
\begin{eqnarray}
c_i(k_i)&\approx&\frac{(pm-1)p^2}{8}\frac{(lnN)^2}{N}+2(1-p)\frac{1}{k_i}\nonumber\\
&\approx&\frac{2(1-p)}{k_i}+\frac{(pm-1)p^2}{8}\frac{(lnN)^2}{N}
\end{eqnarray}

This implies that $\exists k^*$ satisfied that:
\begin{equation}
k^*\approx \frac{16(1-p)N}{(pm-1)p^2\cdot (lnN)^2}
\end{equation}
Then
\begin{equation}
c^*\approx{\frac{(pm-1)p^2(lnN)^2}{4N}}
\end{equation}

So the general solution of the clustering coefficient has the form
$c(k)=\frac{2(1-p)}{k}+\frac{c^*}{2}$. The change rate of the
clustering coefficient $c(k)$ can be tuned by parameter $p$. Fig.3
displays the results of simulation and analytical prediction.

\begin{figure}[th]
\centerline{\psfig{file=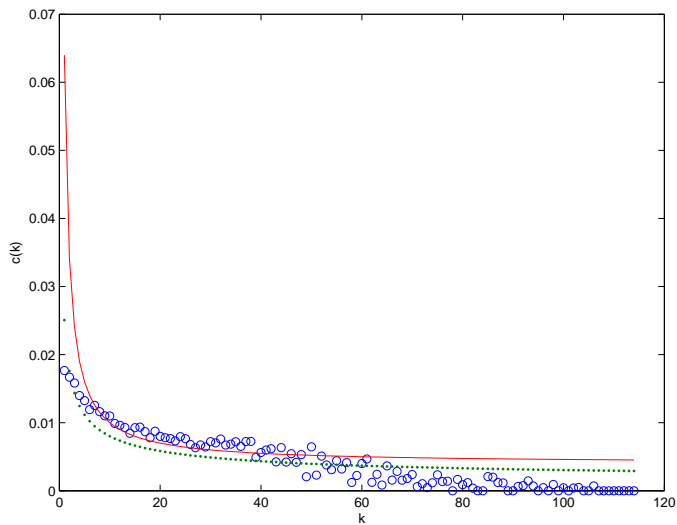,width=9cm}} \vspace*{8pt}
\caption{The clustering coefficient $c$ vs degree $k$ with
$N=10000,m_0=25,M=20,m=3,p=0.97$. Simulation results (o);
Least-square-fit curve (.); Prediction curve (-).
$c(k)=\frac{0.06}{k}+0.004$.}
\end{figure}

\section{Conclusions}
In this paper a developing network with adjustable parameter was
proposed to control the change of the clustering coefficient. The
assumption of this model can be justified by the appearance of
local communities and high transitivity in social networks. With
continuum theory and rate equation method, the analytical
expressions of the degree distribution and the clustering
coefficient have been derived. The analytical predictions are in
good agreement with the numerical results. However, we just made a
first step to explore and understand the evolving mechanisms of
social networks. Many other important factors need to be
considered in the future, such as the left of some persons
periodically. Also for simplicity we just selected the local world
randomly from the network.

\section*{Acknowledgements}
We are grateful to all the anonymous referees who provided
insightful and helpful comments. The work was supported by
NSFC(90103033).

% ------------- References -----------------------------

\end{document}